\documentstyle[sprocl,epsfig]{article}
\begin{document}
\title{Particle Physics I: The SM and the MSSM}
\author{R. D. Peccei}
\address{Department of Physics and Astronomy, UCLA, Los Angeles, CA 
90095-1547}
\maketitle

\begin{abstract}
After discussing alternative
scenarios for the origin of the electroweak symmetry breaking,
I briefly review the experimental status of the Standard Model. I explore further both the hints for, and constraints on, supposing that a supersymmetric extension of the Standard Model exists, with supersymmetry broken at the weak scale. I end with a few comments on the theoretical implications of the recent evidence for neutrino oscillations.
\end{abstract}

It is sad honor to be able to speak at Inner Space/Outer Space II, a
symposium in memory of David Schramm.  Dave was an old friend, whose exuberance
and enthusiasm I greatly miss.  It was from him that I first realized that indeed the cosmos could tell us some things of importance for particle physics.  It is a testament to his influence and vision that now no one doubts that much of what is interesting in high energy physics is writ large in the history of the Universe.  A measure of the changes that have occurred since David first entered the field in the early 1970's is that now cosmological data is often one of the
few weapons that we have to exclude or constrain new ideas in particle physics.  In this respect, Schramm's famous limit on the allowed number of neutrino species, coming from Nucleosynthesis,~\cite{DS} has proven particularly effective as a theory ``sorter"!

\section{Theoretical Issues in the Standard Model}

The Standard Model (SM),~\cite{SM} based on the gauge group $SU(3)\times SU(2)\times U(1)$, has proven very robust, with all precision electroweak
data in excellent agreement with the predictions of the theory.~\cite{Grun}
Nevertheless, there remain important open questions in the SM.  Chief among
them is the mechanism which causes the spontaneous breakdown of
$SU(2)\times U(1)$ to $U(1)_{\rm em}$ and the nature of the symmetry breaking
parameter $v_{\rm F}$---the Fermi scale.
Although the size of $v_F$ $(v_F\sim 250~{\rm GeV})$ is
known, its precise origin is yet unclear.

To understand some of the issues involved, it proves useful to examine the
simplest example of symmetry breakdown in which the symmetry breaking is
effected by just one complex Higgs doublet $\Phi$ in ${\cal{L}}_{\rm SB}$. In this case,
the Fermi scale $v_F$ enters directly as a scale parameter in the
Higgs potential
\begin{equation}
V = \lambda\left[\Phi^\dagger\Phi - \frac{1}{2}v_F^2\right]^2~.
\end{equation}
The sign of the $v_F^2$ term is chosen to
guarantee that $V$ will be asymmetric, with a minimum at a non-zero
value for $\Phi^\dagger\Phi$.  This triggers
the breakdown of $SU(2)\times U(1)$ to $U(1)_{\rm em}$, since it forces $\Phi$ to develop a non-zero VEV:\footnote{With only one Higgs doublet
one can always choose $U(1)_{\rm em}$ as the surviving $U(1)$ in the
breakdown. }
\begin{equation}
\langle\Phi\rangle = \frac{1}{\sqrt{2}} \left(
\begin{array}{c}
v_F \\ 0
\end{array} \right)~.
\end{equation}

Because $v_F$ is an internal scale in the potential $V$, in isolation it clearly makes no sense to ask what physics
fixes the scale of $v_F$.
This question, however, can be asked if one considers the SM as an
effective theory valid up to some very high cut-off scale $\Lambda$, where
new physics comes in. In
this broader context it makes sense to ask what is the relation of
$v_F$ to the cut-off $\Lambda$.  In fact, because the $\lambda\Phi^4$
theory is trivial,~\cite{trivial} with the only consistent theory being
one where $\lambda_{\rm ren} \to 0$, considering the scalar interactions
in ${\cal{L}}_{\rm SB}$ without some high energy cut-off is not sensible.

To appreciate this point, let's compute the evolution of the coupling
constant $\lambda$ from the Renormalization Group equation (RGE)  
\begin{equation}
\frac{d\lambda}{d\ln q^2} = +\frac{3}{4\pi^2}\lambda^2 + \ldots
\end{equation}
This equation, in contrast to what happens in QCD, has a positive rather than a 
negative sign in front of its first term, so $\lambda$ grows with $q^2$.  As a result, if one solves the 
RGE, including only this first term, one finds a singularity at
large $q^2$ which is a reflection of this growth
\begin{equation}
\lambda(q^2) = \frac{\lambda(\Lambda_o^2)}
{1-\frac{3\lambda(\Lambda_o^2)}{4\pi}\ln\frac{q^2}{\Lambda_o^2}}~.
\end{equation}
 Of course,
one cannot trust the location of the, so called, Landau pole ~\cite{Landau} derived from Eq. (3)
\begin{equation}
\Lambda_c^2 = \Lambda_o^2\exp\left[\frac{4\pi^2}{3\lambda(\Lambda_o^2)}\right]~,
\end{equation}
since Eq. (3) stops being valid when $\lambda$ gets too large.  Nevertheless, for any given cut-off $\Lambda_c$,
one {\bf can} predict $\lambda(q^2)$ for scales $q^2$ sufficiently below
this cut-off.  Indeed, the $\lambda\Phi^4$ theory is perfectly sensible
as long as one restricts oneself to $q^2\ll \Lambda_c^2$.  If one wants
to push the cut-off to infinity, however, one sees that
$\lambda(\Lambda_o^2)\to 0$.  This is the statement of triviality,~\cite{trivial}
within this simplified context.

In the case of the SM, one can ``measure" where the cut-off $\Lambda_c$ is in
${\cal{L}}_{\rm SB}$ from the value of the Higgs mass.  Using the
potential (2) one finds that
\begin{equation}
M_H^2 = 2\lambda(M_H^2)v_F^2~.
\end{equation}
Physics is rather different depending on whether the Higgs mass is light or
is heavy with respect to $v_{\rm F}$.  If $M_{\rm H}$ is light 
the effective theory described by ${\cal{L}}_{\rm SB}$ is very
reliable, and {\bf weakly coupled}, with $\lambda\leq 0.3$ up to very high scales.
In these circumstances it is meaningful to ask whether the
large hierarchy $v_F \ll \Lambda_c$
is a stable condition.  This question, following 't Hooft,~\cite{tH}
is often called the problem of {\bf naturalness}.

If, on the other hand, the Higgs mass is heavy, of order of the
cut-off $(M_H \sim \Lambda_c)$, then it is pretty clear that
${\cal{L}}_{\rm SB}$ as an effective theory stops making sense.
The coupling $\lambda$ is so strong that one cannot separate the 
particle-like excitations from the cut-off itself.  Numerical
investigations on the lattice~\cite{lattice} have indicated that this
occurs when
\begin{equation}
M_H \sim\Lambda_c \sim 700~{\rm GeV}~.
\end{equation}
In this case, it is clear that $\langle\Phi\rangle$, as the order
parameter of the symmetry breakdown, must be replaced by something
else.

The Planck scale $M_{\rm P}$ is clearly a natural physical cutoff.  So,
in the weak coupling case, one has to worry whether the hierarchy
$v_{\rm F} \ll M_{\rm P}$ is stable.  It turns out that this is
{\bf not} the case, since radiative effects in a theory with a cutoff destabilize
any pre-existing hierarchy. Indeed,
this was 't Hooft's original argument.~\cite{tH}  Quantities like the Higgs mass that are
{\bf not protected} by symmetries suffer quadratic mass shifts. Schematically, the Higgs mass shifts from the 
value given in Eq. (6) to
\begin{equation}
M_H^2 = 2\lambda v_F^2 + \alpha \Lambda_c^2~.
\end{equation}
It follows from Eq. (8) that
if $\Lambda_c\sim M_P \gg v_F$, the Higgs boson cannot remain
light.  If one wants the Higgs to remain
light one is invited to look for some
protective symmetry to guarantee that the hierarchy $v_F\ll M_P$
is stable.\footnote{Note that a stable hierarchy 
$v_F \ll M_P$ does not explain why such a hierarchy exists.}

Such a protective symmetry exists---it is supersymmetry (SUSY).~\cite{SUSY}
SUSY is a boson-fermion symmetry in which bosonic degrees of freedom
are paired with fermionic degrees of freedom.  If supersymmetry is
exact then the masses of the fermions and of their bosonic partners
are the same.  In a supersymmetric version of the Standard Model
all quadratic divergences cancel.  Thus parameters like the Higgs boson mass
will not be sensitive to a high energy cut-off.  Via
supersymmetry the Higgs boson mass is kept light since its
fermionic partner has a mass protected by a chiral symmetry.

Because one has not seen any of the SUSY partners of the states in the
SM yet, it is clear that if a supersymmetric extension of the SM exists
then the associated supersymmetry must be broken.  Remarkably, even if
SUSY is broken the naturalness problem in the SM is resolved, provided
that the splitting between the fermion-boson SUSY partners is itself
of $O(v_F)$.  For example, the quadratic divergence of the Higgs mass
due to a $W$-loop is moderated into only a logarithmic divergence by the
presence of a loop of Winos, the spin-1/2 partners of the $W$
bosons.  Schematically, in the SUSY case, Eq. (8) gets replaced by
\begin{equation}
M_H^2 = 2\lambda v_F^2 + \alpha(\tilde M_W^2-M_W^2)
\ln \Lambda_c/v_F~.
\end{equation}
So, as long as the masses of the SUSY partners (denoted by a tilde)
are themselves not split away by much more than $v_F$, radiative
corrections will not destabilize the hierarchy $v_F\ll\Lambda_c$.

Let me recapitulate.  Theoretical considerations regarding the nature
of the Fermi scale have suggested two alternatives for new
physics associated with the $SU(2)\times U(1)\to U(1)_{\rm em}$ 
breakdown:
\begin{description}
\item{i)} ${\cal{L}}_{\rm SB}$ is the Lagrangian of some elementary scalar
fields interacting together via an asymmetric potential, whose minimum
is set by the Fermi scale $v_F$.  The presence of non-vanishing
VEVs triggers the electroweak breakdown.  However, to guarantee the
naturalness of the hierarchy $v_F\ll M_P$, both ${\cal{L}}_{\rm SB}$
and the whole SM Lagrangian must be augmented by other
fields and interactions so as to be (at least
approximately) supersymmetric.  Obviously, if this alternative is
true, there is plenty of new physics to be discovered,
since all particles have
superpartners of mass $\tilde m \simeq m + O(v_F)$.
\item{ii)} The symmetry breaking sector of the SM has itself a dynamical
cut-off of $O(v_F)$.  In this case, it makes no sense to describe
${\cal{L}}_{\rm SB}$ in terms of strongly coupled scalar fields.
Rather, ${\cal{L}}_{\rm SB}$ describes a dynamical theory of some new
strongly interacting fermions $F$, whose condensates cause the
$SU(2)\times U(1)\to U(1)_{\rm em}$ breakdown.  The strong interactions
which form the condensates $\langle\bar FF\rangle\sim v_F^3$ also
identify the Fermi scale as the dynamical scale of the underlying
theory, very much analogous to $\Lambda_{\rm QCD}$.  If this alternative
turns out to be true, then one expects also to see lots of new
physics, connected with these new strong interactions, when one probes
them at energies of $O(v_F)$.
\end{description}

\section{Experimental Tests of the SM }

The expectations of the SM, assuming the simplest form of symmetry breaking,
have been confronted experimentally to high accuracy.  These results provide
already some important indications on the nature of the electroweak symmetry
breakdown, which I review here. In
practice, since all fermions but the top are quite light compared to the
scale of the $W$ and $Z$-bosons, all quantities in this simplest version of the  SM are specified as functions
of 5 parameters: $g^\prime$, $g_2$, $v_F$, $M_H$ and $m_t$.  It proves 
convenient to trade the first three of these for another triplet of 
quantities which are better measured:  $\alpha$, $M_Z$ and
$G_F$.  Once
one has adopted a set of {\bf standard parameters} then all physical
measurable quantities can be expressed as a function of this ``standard set".
Because $\alpha$, $M_Z$, and $G_F$, as well as $m_t$\footnote{The top mass
is quite accurately determined now.  The combined value obtained by the CDF
and DO collaborations fixes $m_t$ to better than 3\%:  $m_t =
(173.8 \pm 5.0)$ GeV.~\cite{Tevatron}} are rather accurately known, all SM
fits essentially constrain only {\bf one} unknown-- the Higgs mass
$M_H$.  This constraint, however, is not particularly strong because all
radiative effects depends on $M_H$ only logarithmically.

The result of the SM fit of all precision data gives for the Higgs mass ~\cite{CPSLAC1}
\begin{equation}
M_{\rm H} = \left(98^{+ 57}_{- 38} \right)~{\rm GeV}
\end{equation}
and the 95\% C.L. upper bound: $M_{\rm H} < 235~{\rm GeV}$.  It is particularly
gratifying that this fit indicates the need for a light Higgs boson, since
this ``solution" is what is internally consistent.  Furthermore, this
result is also compatible with the limit on $M_{\rm H}$
coming from direct searches for the Higgs boson in the process
$e^+e^-\to ZH$ at LEP 200.  The limit given at the 1999 Lepton Photon Conference
at SLAC is,~\cite{LPSLAC2} at 95\% C.L.,
\begin{equation}
M_{\rm H} > 95.2~{\rm GeV}~.
\end{equation}
By running LEP 200 at $\sqrt{s} = 200$ GeV in the coming year one opens up
another 10 GeV of discovery potential for the Higgs boson.  The present
Tevatron bounds for $M_{\rm H}$ are weaker, being roughly a factor of 20-50
too insensitive for $M_{\rm H} = 95$ GeV.  However, with the substantial
luminosity increased planned, the
Tevatron can explore a Higgs window up to $M_{\rm H}$ = 110-130 GeV,~\cite{TeVHiggs} before the turn-on of the LHC. The LHC, of course, has the capability of exploring the full range for $M_H$, well beyond the upper bound estimate (7).

The physical lower bound (11) suffices to rule out the possibility of electroweak baryogenesis within the context of this
simplest version of the SM.  To allow for electroweak baryogenesis it is
necessary that the $SU(2)\times U(1)$ phase transition be strongly first order.~\cite{Shap}  Only in this case can one prevent having the (B+L)-violating interactions in the SM going back into equilibrium after the electroweak phase transition,
thereby erasing any matter asymmetry established during the phase
transition.  One can show that to prevent erasing the established asymmetry
one needs the order parameter at the phase transition to have a value
$\langle\Phi(T^*)\rangle/T^*\geq 1$.  Such a large jump in the Higgs VEV, however, only
occurs for relatively light Higgs boson masses---typically $M_{\rm H}
\stackrel{<}{_{\scriptstyle \sim}}$ 50 GeV,~\cite{EWbaryo} with
$\langle\Phi(T^*)\rangle/T^*$ decreasing rather rapidly as $M_{\rm H}$
increases.\footnote{In fact, for the minimal SM, the matter asymmetry 
established at the electroweak phase transition (before its erasure) is also
much below what is needed because there is not enough CP violation, due to
GIM suppression factors.~\cite{HN}}

Within the context of this simplest version of the SM one expects
$M_{\rm H}$ to be larger than the bound (11) from the requirement of vacuum stability.
The argument is rather simple.  Because top is rather
heavy, in the RGE for the Higgs coupling $\lambda$ one cannot neglect the
effect of the top Yukawa coupling.  Thus, instead of Eq. (3) one has
\begin{equation}
\frac{d\lambda}{d\ln \mu^2} = \frac{3}{4\pi^2}
\left[\lambda^2 - \frac{1}{4}\lambda_t^4\right] + \ldots
\end{equation}
Because the top contribution comes with a {\bf negative} sign, it will slow
down and can actually reverse the growth of $\lambda$. Indeed, if the Higgs coupling $\lambda(M_H)$ is
not large enough, because the Higgs boson is light, the 
contribution coming from the $\lambda_t^4$ term can drive $\lambda$
{\bf negative} at some scale $\mu$.  This cannot happen physically,
because for $\lambda < 0$ the Higgs potential is unbounded!

To avoid this vacuum instability below some cut-off $\Lambda_c$ one needs to have $\lambda(M_H)$, and therefore
the Higgs mass, sufficiently large.  Hence, these considerations 
give a lower bound for the
Higgs mass.  Taking $\Lambda_c = M_{\rm P}$, this 
lower bound is~\cite{lowerbound}
\begin{equation}
M_H \geq 134~{\rm GeV}~.
\end{equation}
Lowering the cut-off $\Lambda_c$, weakens the bound
on $M_H$.  Interestingly, to have a SM Higgs as light as 100 GeV-- which
is the region accessible to LEP 200 and the Tevatron-- requires a very low cut-off,
of order $\Lambda_c\sim 100$ TeV.~\cite{Lindner}

Of course, a good fit  of the data with the minimal SM does not necessarily exclude possible
extensions of the SM involving either new particles or new interactions,
provided that these new particles and/or interactions give only small effects.
Typically, the effects of new physics are small if the excitations associated 
with this new physics have mass scales several times the $W$-mass.
One can quantify the above discussion in a more precise way by introducing
a general parametrization for the vacuum polarization tensors of the gauge
bosons and the $Zb\bar b$ vertex. These are the places where the dominant
electroweak radiative corrections occur and therefore are the quantities
which are most sensitive to new physics.~\cite{epsilon}  As an illustration, I will discuss an example which has a bearing on the nature of the electroweak symmetry breaking.

There are four distinct vacuum polarization contributions
$\Sigma_{AB}(q^2)$, where the pairs $AB = \{ZZ,WW,\gamma\gamma,$ $
\gamma Z\}$.  For sufficiently low values of the momentum transfer
$q^2~(q^2\simeq M^2_W)$ it obviously suffices to expand 
$\Sigma_{AB}(q^2)$ only up to $O(q^2)$.  Thus, approximately,
one needs to consider 8 different parameters associated with these
contributions:
\begin{equation}
\Sigma_{AB}(q^2) = \Sigma_{AB}(0) + q^2\Sigma_{AB}^\prime(0) + \ldots~.
\end{equation}
 In fact, there are not really 8 independent parameters since
electromagnetic gauge invariance requires that
$\Sigma_{\gamma\gamma}(0) = \Sigma_{\gamma Z}(0) = 0$.
Of the 6 remaining parameters one can fix 3 combinations of coefficients in
terms of $G_F,~\alpha$ and $M_Z$.  Hence, in a most general analysis, the
gauge field vacuum polarization tensors (for $q^2\stackrel{<}{_{\scriptstyle
\sim}} M_W^2$) only involve 3
arbitrary parameters.  The usual choice,\cite{epsilon} is to have one of
these contain the main quadratic $m_t$--dependence, leaving the other two
essentially independent of $m_t$.
In the notation of Altarelli and Barbieri, the parameter that depends on
$m_t$ is  called $\epsilon_1$, with $\epsilon_2$ and $\epsilon_3$ being at most
logarithmically dependent on this mass.
For our purpose, the interesting parameter is $\epsilon_3$, whose
value, obtained from a fit of all precision electroweak data, turns out to be \cite{ABC}
\begin{equation}
\epsilon_3 = (3.9 \pm 1.1) \times 10^{-3}~.
\end{equation}

Given some assumption of how $SU(2)\times U(1)$ is broken down, one can
estimate the various $\epsilon_i$ parameters.  This is somewhat harder to do
in theories where the spontaneous breakdown occurs dynamically,  since these
involve strong interactions in the symmetry breaking sector.  Nevertheless,
one can estimate $\epsilon_3$ in a dynamical symmetry breaking theory,
if one assumes that the spectrum of such a theory, and its dynamics, is
QCD-like.\cite{Peskin}  From its definition $\epsilon_3$
involves the difference between the spectral functions of vector and
axial vector currents
This difference has two components in a dynamical symmetry breaking theory.
There is a contribution from a heavy Higgs boson $(M_H\sim {\rm TeV})$
characteristic of such theories, plus a term detailing the differences between the vector
and axial vector spectral functions. 
This second component reflects the spectrum of the 
underlying theory which causes the symmetry
breakdown.  The first piece is readily computed from the SM expression,
using $M_H\sim {\rm TeV}$.  The second piece, in a QCD-like theory, can
be estimated, modulo some counting factors.  One finds~\cite{Peskin}
\begin{equation}
\epsilon_3 =\left[6.65\pm 3.4 N_D\left(\frac{N_{TC}}{4}\right)\right]
\times 10^{-3}~.
\end{equation}
The second term follows if the underlying theory is QCD-like, so that
the resonance spectrum is saturated by $\rho$-like and $A_1$-like, resonances.
Here $N_D$ is the number of doublets entering in the underlying theory
and $N_{TC}$ is the number of ``Technicolors" in this theory.\footnote{For
QCD, of course, $N_D=1$ and $N_{TC}=3$.}  Using Eq. (16) and taking
$N_{TC}=4$, as is usually assumed, one sees that
\begin{equation}
\epsilon_3 = \left\{ \begin{array}{ll}
10.05\times 10^{-3} & N_D=1 \\
20.25\times 10^{-3} & N_D=4
\end{array}
\right.
\end{equation}
These values for $\epsilon_3$ are, respectively, 
$5.5\sigma$ and $15\sigma$ away from the best fit value of $\epsilon_3$.  Obviously,
one cannot countenance anymore a dynamical symmetry breaking theory which
is QCD-like!

Nothing as disastrous occurs instead if one considers a supersymmetric
extension of the SM, provided the superpartners are not too light.  
Fig. 1, taken from a recent analysis of Altarelli, Barbieri, and Caravaglios,\cite{ABC} shows a typical fit, scanning over a range of
parameters in the MSSM---the minimal supersymmetric extension of the SM.
Although the MSSM improves the $\chi^2$ of the fit over that for
the SM (which is already very good!), these improvements are small and one cannot use this as evidence for low energy supersymmetry.

\begin{figure}[t]
\center
\epsfig{file=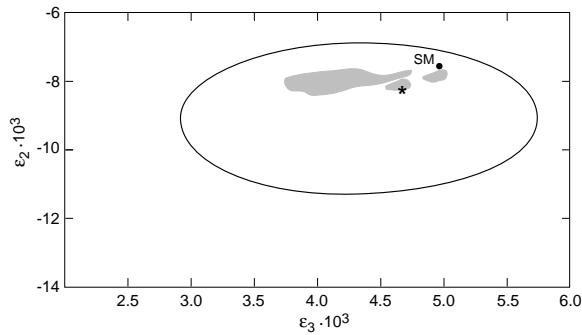,height=4in}
\caption{Comparison of SM and MSSM fits in the $\epsilon_2-\epsilon_3$ plane, from Ref.19. The ellipse is the 1-$\sigma$ range determined by the data. The shaded region is the result of a scan over a range of SUSY parameters, with the star marking the lowest $\chi^2$ point.}
\end{figure}

\section{Searching for Supersymmetry}
\subsection{Unification of Couplings}

An indirect piece of evidence favoring the existence of supersymmetry at the
weak scale is the way the $SU(3)\times SU(2) \times U(1)$ coupling constants evolve with
energy.  Although these couplings are quite different at low energy,
 they evolve differently with $q^2$.  In
leading order, the RGE 
\begin{equation}
\frac{d\alpha_i(\mu^2)}{d\ln\mu^2} = -\frac{b_i}{4\pi}\alpha_i^2(\mu^2)~.
\end{equation}
 imply a logarithmic change for the inverse couplings.
The rate of change of the coupling constants with energy is governed by the
coefficients $b_i$ which enter in the RGE.  In turn, these coefficients
depend on the {\bf matter content} of the theory---which 
matter states are ``active" at the 
scale one is probing.
Remarkably, with ordinary matter, one gets near {\bf unification} of couplings
at high energy.  However, assuming that there are supersymmetric partners
of ordinary matter present above the Fermi scale, the three SM couplings
really unify, as shown in Fig. 2!

\begin{figure}[t]
\center
\epsfig{file=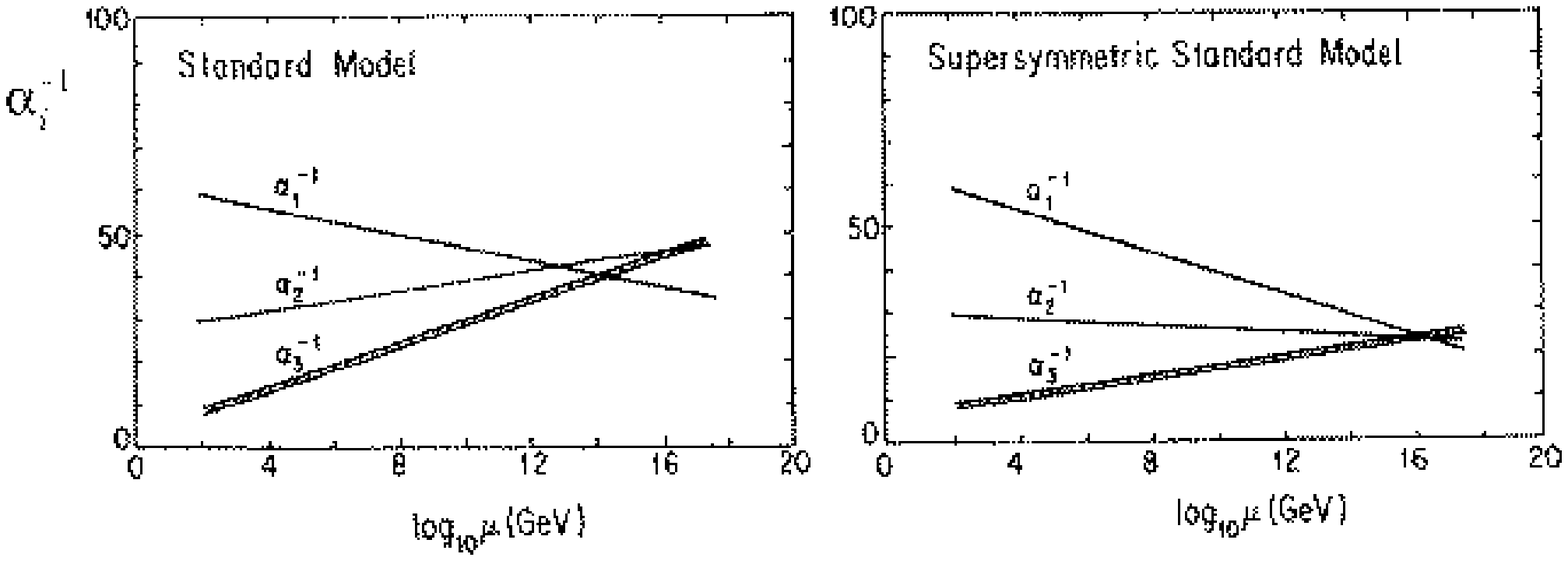,width=0.9 \textwidth}
\caption{Evolution of couplings without and with SUSY matter.}
\end{figure}

The unification of the couplings in the SUSY SM case is quite spectacular.
However, {\it per se}, this is only suggestive.  It is neither a
``proof" that a low energy supersymmetry exists, nor does it mean that there exists
some high energy
Grand Unified Theory (GUT) which breaks down to the SM at a high scale. The proof of the former requires the discovery
of the predicted SUSY partners, while for GUTs one must find typical
phenomena which are associated with these theories--like proton decay.
Nevertheless, if such a GUT exists, one learns that the unification scale is rather high
in the supersymmetric case [$M_X \simeq 2\times 10^{16}$ GeV].  Such high
scales gives unobservedly long lifetimes for proton decay arising from
$d=6$ $qqq\ell$ operators, since this lifetime scales as $\tau_p \sim
M_X^4$.  In SUSY GUTs, however, one has dangerous $d=5$ operators where two
quarks are replaced by squarks resulting in a $\tilde q\tilde qq\ell$ operator.  These terms
lead to rather rapid proton decay, unless they are suitably suppressed.  As a
result, the predictions of SUSY GUT models for proton decay are rather model-dependent, with
modes involving strangeness in the final state, like $p\to \nu K$, dominating.\cite{AN}  For these reasons the existing bounds on proton decay~\cite{KanB} only serve to constrain parameters and cannot be 
adduced either for or against supersymmetry.

Much the same comments can be made regarding more
careful calculations of the evolution of the SM couplings, including both
2-loop effects and more detailed SUSY thresholds.  It turns out that the results of these more refined calculations, assuming a high scale unification of
couplings, do not quite give the correct value of $\alpha_3(M^2_Z)$, unless one
assumes a rather high average SUSY threshold $T_{\rm SUSY}\sim 1$ TeV.~\cite{2loop}  However, this is probably not a serious problem, since
one could well imagine being able to lower $T_{\rm SUSY}$ as a result of a few
percent correction to the evolution equations coming from, difficult to pin 
down, GUT thresholds.~\cite{GUTth}

\subsection{SUSY Higgs Sector}

Supersymmetry associates bosonic partners to fermions and {\it vice versa}.
However, it also requires two Higgs doublets, since the superpotential
which describes a SUSY extension of the Yukawa interactions in the SM can
only contain chiral superfields and not their adjoint.
Although $H_u^*$ has the same quantum number as $H_d$, supersymmetry does not
allow this more parsimonious choice.  As a result, all supersymmetric
extensions of the SM necessarily imply the presence of 5 physical Higgs
states.  Three of these states are neutral $(h,H,A)$ and two are charged
$(H^\pm)$, with $h$ and $H$ being scalar and $A$ pseudoscalar.

The minimal set of Higgs states which appear in a supersymmetric extension of
the SM has another remarkable property.  Their quartic interactions are
entirely {\bf fixed} by supersymmetry, since they arise from the structure
of the gauge interactions dictated by supersymmetry---the, so called, $D$-terms.~\cite{SUSY}  No other quartic terms can be induced by supersymmetry
breaking, if one wants to have supersymmetry be the solution to the 
hierarchy problem, since such $d=4$ terms would trigger a hard breaking of
supersymmetry.  However, supersymmetry breaking can affect the $d=2$ terms in the Higgs potential.  As a result, in this minimal supersymmetric extension
of the SM---the, so called, MSSM---one can write down a quite specific Higgs
potential

\begin{eqnarray}
V(H_u,H_d) &=& (H_u^\dagger~ H_d^\dagger){\cal{M}}^2 \left(
\begin{array}{c}
H_u \\ H_d
\end{array} \right) +
\frac{1}{8} g^{\prime 2}[H_u^\dagger H_u-H_d^\dagger H_d]^2 \nonumber \\
& & +\frac{1}{8} g_2^2[H_u^\dagger\vec\tau H_u + H_d^\dagger
\vec\tau H_d]\cdot[H_u^\dagger\vec\tau H_u + H_d^\dagger\vec\tau H_d]~.
\end{eqnarray}
Here the mass squared ${\cal{M}}^2$ contains both SUSY preserving terms ($\mu$) and SUSY breaking terms ($ B$ and $\mu_{ij}$):
\begin{equation}
{\cal{M}}^2 = \left(
\begin{array}{cc}
\mu^2 + \mu_{11}^2 & -B\mu + \mu_{12}^2 \\
-B\mu + \mu_{12}^2 & \mu^2 + \mu^2_{22} 
\end{array} \right) \equiv \left(
\begin{array}{cc}
m_1^2 & m_3^2 \\ m_3^2 & m_2^2 
\end{array} \right)~.
\end{equation}
Obviously, a breakdown of $SU(2)\times U(1)\to U(1)_{\rm em}$ requires that
${\rm det}~{\cal{M}}^2 < 0 $.

The Higgs mass spectrum arising from Eq. (19) can be parametrized as a
function of one of the masses, usually taken to be $M_A$, and the ratio of the
two Higgs fields VEVs: $\tan\beta = \langle H_u\rangle/\langle H_d\rangle$.
One finds~\cite{Higgsguide}
\begin{eqnarray}
M^2_{H^\pm} &=& M^2_A + M^2_W \nonumber \\
M^2_{H,h} &=& \frac{1}{2} (M_A^2 + M^2_Z) \pm \frac{1}{2}
\left[(M_A^2 + M^2_Z)^2 - 4M_Z^2 M_A^2\cos^22\beta\right]^{1/2}~.
\end{eqnarray}
It is easy to see from Eq. (21) that there is always one {\bf light Higgs}
in the spectrum:
\begin{equation}
M_h \leq M_Z|\cos 2\beta| \leq M_Z~.
\end{equation}
However, the bound of Eq. (22) is not trustworthy, as it is quite sensitive
to radiative effects which are enhanced by the large top mass.~\cite{Higgsred}
Fortunately, the magnitude of the radiative shifts for $M^2_h$ can be well
estimated, by either direct computation~\cite{Hollik} or via the
renormalization group.~\cite{Haber}

As an example, in the case where $M_A\to\infty$ and $|\cos 2\beta|\to 1$, one
finds~\cite{Haber1}
\begin{equation}
M_h^2 = M^2_Z + \frac{3\alpha}{2\pi\sin^2\theta_W}
\left[\frac{m_t^4}{M^2_W}\right] \ln(M^2_{\rm SUSY}/M^2_Z)~,
\end{equation}
where $M_{\rm SUSY}$ is an assumed common scale for all the SUSY partners.
As one can appreciate from the above formula, this is quite a large
shift since, for 
$M_{\rm SUSY}\simeq 1$ TeV, one finds
$\Delta M_h\simeq 20$ GeV.
Eq. (23) was obtained in a particular limit $(|\cos 2\beta|\to 1)$, but
an analogous result can be obtained for all $\tan\beta$. 
For small $\tan\beta$
the shifts are even larger than those indicated in Eq. (23). However, for these values of $\tan\beta$ the tree level contribution is also
smaller, since $\left.M_h\right|_{\rm tree} < M_Z\cos 2\beta$.  
The actual details of the SUSY spectrum are in general not very important. The biggest effect of
the SUSY spectrum for $\Delta M_h$ 
arises if there is an incomplete cancellation between
the top and the stop contributions, due to large $\tilde t_{\rm L}-\tilde t_{\rm R}$ mixing.~\cite{Haber1}  At their maximum these effects
cause a further shift of order $(\Delta M_h)_{\rm mixing} \simeq 10$ GeV.

One can contrast these predictions with experiment.  At LEP 200,
the four LEP collaborations have looked both for the process 
$e^+e^-\to hZ$ and $e^+e^-\to hA$.  The first process is analogous to that
used for searching for the SM Higgs, while $hA$ production is peculiar
to models with two (or more) Higgs doublets.  One can show that these two
processes are complementary, with one dominating in a region of parameter 
space where the other is small, and {\it vice versa}.~\cite{Higgsguide}
LEP 200 has established already rather strong bounds for $M_h$ and $M_A$
setting the 95\% C.L. bounds (for $\tan\beta > 0.4$)~\cite{LPSLAC2}
\begin{equation}
M_h > 80.7~{\rm GeV}~; ~~~~ M_A > 80.9~{\rm GeV}
\end{equation}
As Fig. 3 shows, if there is not much $\tilde t_{\rm L}-\tilde t_{\rm R}$
mixing the low $\tan\beta$ region $[0.8 < \tan\beta < 2.1]$ is also already
excluded.

\begin{figure}
\begin{center}
\epsfig{file=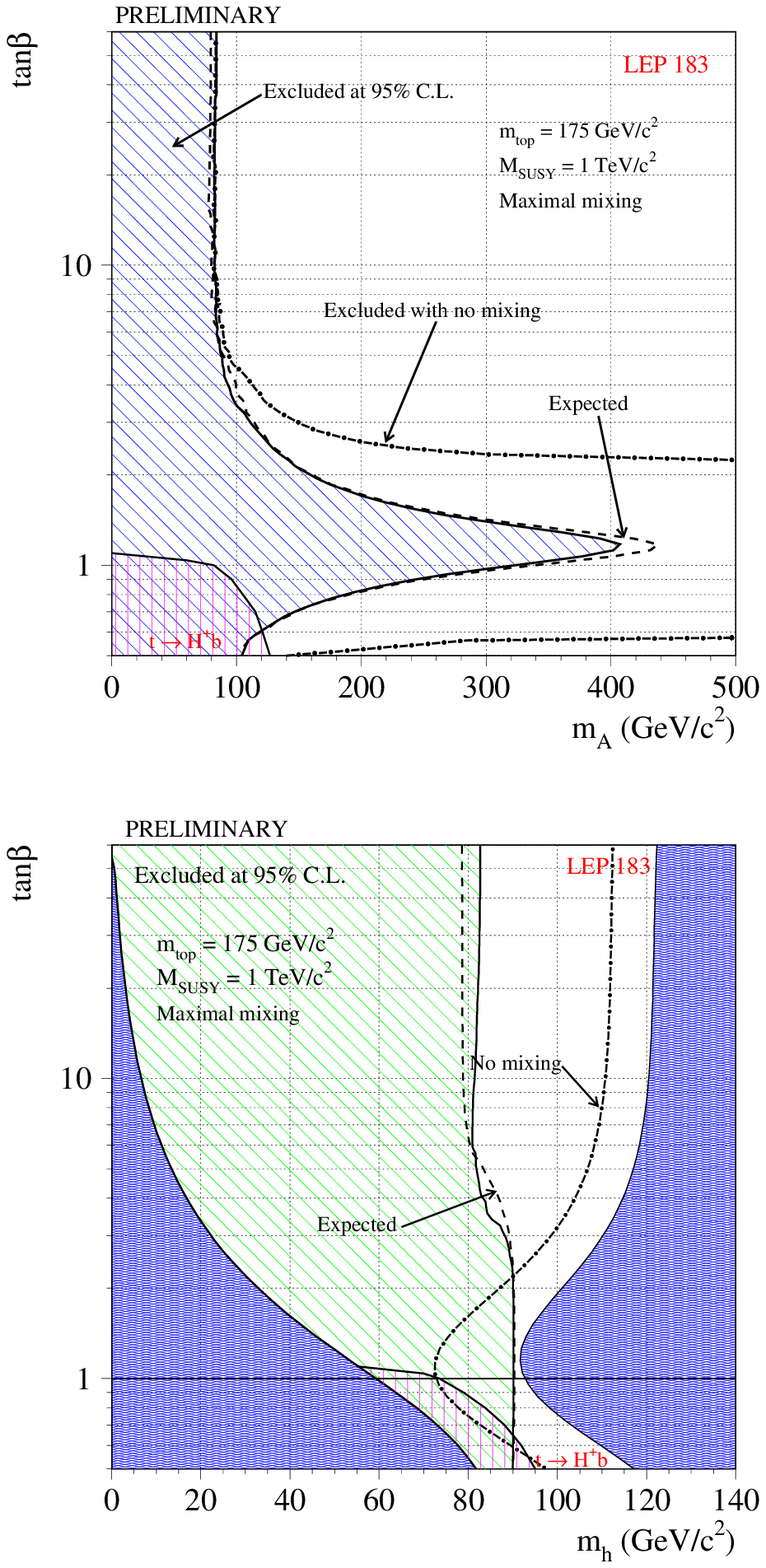,height=6in}
\end{center}
\caption{LEP200 limits for$M_h$ and $M_A$ as a function of $\tan \beta$, from Ref. 30.}
\end{figure}

It is apparent that the available window for $M_h$ is tantalizing small even
for larger $\tan\beta$.  LEP 200, running at its maximum energy of 200 GeV
and the upgraded Tevatron with more luminosity can explore a good deal more still,
providing even more stringent tests for the MSSM.  In fact, because more
complicated supersymmetric extensions of the SM (e.g. those obtained by including additional
gauge singlet Higgs superfields~\cite{singlet}) retain the same
qualitative features, probing in detail the Higgs spectrum is a very effective
way to test the whole notion of the existence of an approximate supersymmetry
at the weak scale.

\subsection{Sparticle Searches}

Although the SUSY SM is rather predictive when it comes to the Higgs sector,
beyond this sector the spectrum of SUSY partners and possible allowed
interactions is quite model dependent.  Most supersymmetric extensions of the
SM considered are assumed to contain a discrete symmetry, $R$-parity, which
is conserved.  This assumption restricts the form of the
possible interactions allowed.  In fact, $R$-parity
conservation provides an essentially unique way to generalize the SM since
$R$, defined by $R = (-1)^{Q+L+2J}$,
with $Q$ being the quark number, $L$ the lepton number and $J$ the spin,
turns out simply to be +1 for all particles and -1 for all sparticles.

Obviously, $R$ parity conservation implies that SUSY particles enter in
vertices always in pairs, and hence sparticles are always pair-produced.
This last fact implies, in turn, the {\bf stability} of the lightest
supersymmetric particle (LSP), even in the presence of supersymmetry
breaking interactions, since these SUSY
breaking interactions can be arranged to preserve R-parity. 
Because the LSP has interactions of weak scale strength, this particle is an excellent dark matter candidate.  This point is discussed in much more
detail by John Ellis in this conference.~\cite{Ellis}

The decay chains of sparticles is predicated both on the modality of
supersymmetric symmetry breaking and on the nature of the LSP.  Of course,
these two are intimately connected. Let me discuss the issue of SUSY breaking in a little more detail, since the manner in which one breaks supersymmetry is the
principal source of model-dependence for the SUSY SM.
In general,~\cite{Peskin2} one assumes that SUSY is spontaneously broken at
some scale $\Lambda$ in some {\bf hidden sector} of the theory.  This sector is
coupled to ordinary matter by some {\bf messenger} states of mass $M$, with
$M \gg \Lambda$, and all that obtains in the visible sector is a set of soft
SUSY breaking terms---terms of dimension $d<4$ in the Lagrangian of the 
theory.\footnote{Recall that terms of $d=4$ would re-introduce the hierarchy problem.}
Ordinary matter contains supersymmetric states with masses $\tilde m\sim$ TeV,
with $\tilde m$ given generically by
$\tilde m\sim \Lambda^2/M.$

Within this general framework, two distinct scenarios have been suggested
which differ by what one assumes are the messengers that connect the hidden
SUSY breaking sector with the visible sector.  In supergravity models~\cite{SUGRA} (SUGRA), the messengers are gravitational interactions,
so that $M\sim M_P$.  Then the demand
that $\tilde m\sim$ TeV fixes the scale of SUSY breaking in the hidden sector to be of order $\Lambda\sim 10^{11}$ GeV.  In contrast, in models where the
messengers are gauge interactions ~\cite{GMM} (Gauge Mediated Models) with
$M\sim 10^6$ TeV, then the scale of spontaneous breaking of supersymmetry is
around $\Lambda\sim 10^3$ TeV.

In both cases one assumes that the supersymmetry is a local symmetry, gauged
by gravity.~\cite{Ferrara}  Then the massless fermion which originates from
spontaneous SUSY breaking, the goldstino, is absorbed and serves to give
mass to the spin-3/2 gravitino---the SUSY partner of the graviton,
whose mass is of order $
m_{3/2} \sim \Lambda^2/M_{\rm P}.$
Obviously, in SUGRA models the gravitino has a mass of the same order as
all the other SUSY partners ($\tilde m \sim$ TeV),  
but there is no reason why 
the gravitino should be the LSP.  However, in Gauge
Mediated Models, since $\Lambda \ll 10^{11}$ GeV, the gravitino
is definitely the LSP.

Besides the above difference, the other principal difference between SUGRA
and Gauge Mediated Models of supersymmetry breaking is the assumed form of
the soft breaking terms.  In SUGRA models, to avoid FCNC problems, one
needs to assume that the soft breaking terms are {\bf universal}.  This
assumption is unnecessary in Gauge Mediated Models, where in fact one can
explicitly compute the form of the soft breaking terms and show that they do not
lead to FCNC.

The search strategies and the resulting bounds on sparticles are quite model-dependent.  For instance, in Gauge Mediated Models, in general the
lightest chargino and neutralino have related masses
($m_{\tilde{\chi}_1^+}\simeq 2m_{\tilde{\chi}_1^o}$).~\cite{Giudice}  Thus the lightest
chargino decays always to the lowest neutralino and this state in turn
radiatively decays to the gravitino LSP.  At the Tevatron, therefore, if one
produces charginos, these would typically produce decays with two photons
and missing energy from the chain $\tilde\chi_1^+\to W^+\tilde\chi_1^o\to
W^+\gamma\tilde G$.

Unfortunately, no signals of SUSY states have been found yet.  To
illustrate the nature of the present bounds, I will describe these bounds for the MSSM, when the SUSY symmetry breaking is
assumed to have a SUGRA origin. This scenario
is characterized by just a few universal
parameters, which include a common scalar mass $m_o$ and a universal gaugino mass $m_{1/2}$, at scales of order $M_X$.
Given these inputs, one can then derive the spectrum of the SUSY states,
essentially by using the RGE evolution from $M_X$ to the weak scale.  In this
model at low energy the ratio of gaugino masses follows the
ratio of the gauge coupling constants, so that
\begin{equation}
M_1:M_2:M_3 = \alpha_1:\alpha_2:\alpha_3 = 1:2.45:8.62~.
\end{equation}
This pattern is quite different from that which obtains if the supersymmetry
breaking at large scales is mediated by scale anomalies, as suggested
recently by Randall and Sundrum.\cite{RS}  In this latter case 
$M_i = -b_i\alpha_iM_{\rm SUSY}$ and thus
\begin{equation}
M_1:M_2:M_3 = 3.3:1:-10~.
\end{equation}
Obviously the phenomenology is  also quite different, since now
$ m_{\tilde{\chi}_1^+} \simeq m_{\tilde{\chi}_1^o}$.

For the MSSM with the simplest universal SUGRA breaking, not surprisingly
the best bounds for the strongly interacting sparticles [squarks, $\tilde q$,
and gluinos, $\tilde g$] come from the Tevatron, while LEP 200 gives the best
bounds for weakly interacting sparticles [sleptons, $\tilde\ell$, and weak
gauginos, both $\tilde\chi_1^\pm$ and $\tilde\chi_1^o$].  Typically, for $m_{\tilde g} \sim m_{\tilde q}$ the mass limits are above 250 GeV for squarks and gluinos~\cite{gluinos}. For the sleptons, the LEP limits are near half the CM energy, with the most recent analysis giving, at the 95\% C.L.~\cite{LPSLAC2}  
 \begin{equation}
m_{\tilde e}> 89 {\rm GeV}~~;m_{\tilde \mu}>81 {\rm GeV}~~;m_{\tilde \tau}> 71 {\rm GeV}.
\end{equation}

The stop limits are a special case.  Because of the large top mass there can be
sizable $\tilde t_{\rm L}-\tilde t_{\rm R}$ mixing, so that the stop eigenvalues can have large splittings.  The lightest stop, $\tilde t_1$, can be searched for at LEP 200, as well as at the Tevatron where it can either be pair produced,
 or can originate from top decay:
$t\to\tilde t_1\tilde\chi_1^o$. ~\cite{HVan}  The LEP 200 stop bounds are (slightly)
dependent on the $\tilde t_{\rm L}-\tilde t_{\rm R}$ mixing angle.  The
Tevatron bound, on the other hand, is quite strong provided that the LSP is light $(M_{\rm LSP} \leq 50~{\rm GeV})$, and one finds, at
95\% C.L. $
\tilde m_{t_1} > 122~{\rm GeV}~.$
Both this bound, as well as the LEP 200 bounds, are shown in Fig. 4.

\begin{figure}
\begin{center}
\epsfig{file=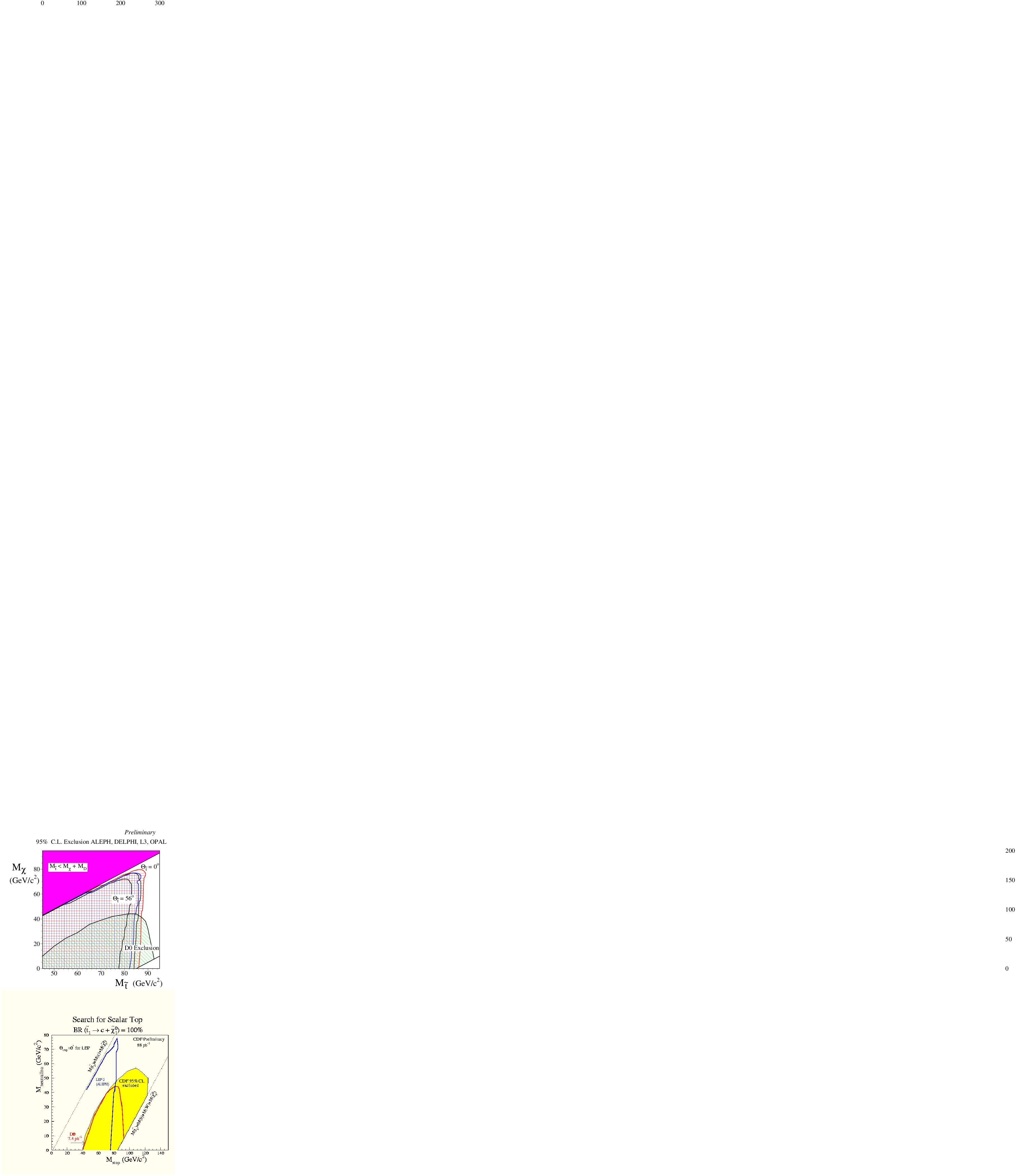,height=6in}
\end{center}
\caption{ Tevatron and LEP200 limits for $ \tilde{m}_t$ , from Ref. 30.}
\end{figure}

\subsection{Electroweak Baryogenesis}

The bounds on $M_h$ and on the lighest stop state are quite relevant to the whole issue of electroweak baryogenesis. This is because   if  one has a rather light stop this helps make the electroweak phase transition more strongly first order, \cite{lightstop} thereby ameliorating the need for having a very light Higgs. One can understand this qualitatively as follows. A light stop modifies significantly the coefficient of the {\bf cubic} term in the Higgs effective potential \footnote{ Here $\phi$ is the effective scalar field describing the electroweak phase transition in the model. It is an open question whether,  for the purposes of discussing the electroweak phase transition, it really suffices to reduce the multi-Higgs SUSY potential in this manner.}
\begin{equation}
V_{eff}= -m^2(T)\phi^2  +\lambda(T) \phi^4 - {\cal E}(T) \phi^3~.
\end{equation}
Because
\begin{equation}
\frac{\langle \phi(T^*)\rangle}{T^*}\simeq \frac{{\cal E}(T^*)}{\sqrt{2}\lambda(T^*)}~,
\end{equation}
if ${\cal E}(T^*)$ is larger this allows $\lambda(T^*)$, and hence the Higgs mass, to be larger for a given jump in the Higgs VEV at the phase transition.

In practice, however, one must be careful with what goes on with the rest of the Higgs potential. In particular, one must make sure that the relaxation of the Higgs bound occurs in a region of parameter space where ther is {\bf no} charge or color breaking minima. Fig. 5, adapted from a paper by  Carena, Quiros and Wagner, ~\cite {CQW}
shows the region in the $M_h-\tilde{m}_t $ plane for which $\frac{\langle \phi(T*)\rangle}{T*}\geq 1.$ Given the present bounds on $M_h$ and $\tilde{m}_t$, there appears to be little phase space still allowed, particularly if one excludes the cosmologically troubling possibility of having a two-step electroweak phase transition. Of couse, this graph is somewhat misleading because it projects all the existing 
parameter space onto the $M_h-\tilde{m}_t$ plane. In addition, one should remember that the $\tilde{m}_t$ bounds where obtained under some assumptions-- e.g. that $m_{\tilde{\chi}_1^0}< $50 GeV and that the lightest stop decayed 100\% of the time via $\tilde{t}_1 \to c  \tilde{\chi}_1^0$. Nevertheless, taking Fig. 5 at face value,  the allowed region is rather small. Furthermore, particularly the large $\tilde{m}_t$ piece slice has other problems, since large $\tilde{m}_t$ goes hand in hand with low $\tan \beta$, which is not favored by the SUSY Higgs searches. Given the above, it is not inconceivable to me that, even before the turn-on of the LHC. one may shut the window for electroweak baryogenesis also in the MSSM.

\begin{figure}[t]
\begin{center}
\epsfig{file=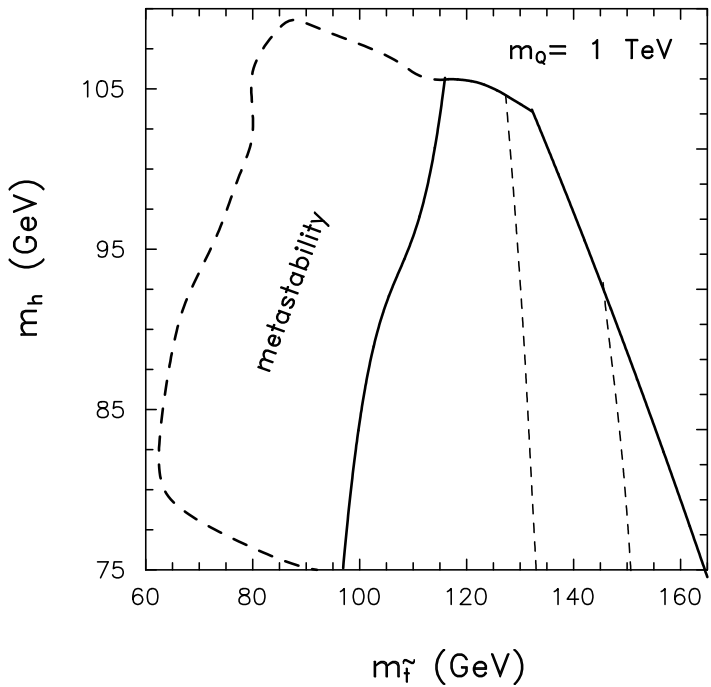,height=4in}
\end{center}
\caption{Allowed region for electroweak baryogenesis in the $M_h-\tilde{m}_t $ plane, from Ref. 41. The short-dashed lines demark the region where a two-step phase transition may occur.}
\end{figure}

Of course, the non erasure of the matter asymmetry produced at the electroweak phase transition is only a {\bf necessary condition} for electroweak baryogenesis. One needs also be able to produce a sufficient asymmetry during the phase transition ($\eta\sim 5 \times 10^{-10}$). The calculation of $\eta$ depends crucially on the strength of the CP violation associated with Baryon-violating processes. In this respect, supersymmetric theories have an advantage, since it is possible to have CP-violating phases which are {\bf flavor conserving}. An example of such a phase is that associated with the parameter $\mu=|\mu|e^{i \alpha_{\mu}}$ in the SUSY-preserving Higgs superpotential $W=\mu H_u H_d$. Another example is the phase entering in the SUSY-breaking gaugino mass $m_{1/2}=|m_{1/2}| e^{i \alpha_{1/2}}$. The calculation of $\eta$ in SUSY models in the literature typically requires that the CP-violating phase which enters in establishing the asymmetry is not too small--typically, $\alpha_{asym}\sim 10^{-1}- 10^{-2} $~\cite{SUSYEW}. It is not clear, however, whether such largish SUSY CP phases are in contradiction with the bounds one obtains on these phases from the neutron dipole moment, $\alpha_{edm}\leq 10^{-2}$ ,~\cite{SUSYEDM} since $\alpha_{asym}$ and $\alpha_{edm}$ are not simply related to each other.

\section{What are Neutrinos Telling Us?}

Precision meausurements of the Z-line shape now determine the number of 
light neutrinos species to a very high accuracy-- much greater that that which can be obtained from Nucleosynthesis. 
The value of $N_{\nu}$ extracted from the combined data of all four LEP experiments, ~\cite{Swa} $N_{\nu} = 2.9835\pm 0.0083~,$ gives very strong evidence that there exist only three families of quarks and leptons.

Although the result on $N_{\nu}$ is remarkable, the most exciting news from neutrinos in the last year is the evidence coming from SuperKamiokande~\cite{SuperK} for neutrino oscillations. In a simple 2-neutrino description, ~\cite{RDP} the SuperKamiokande results are consistent with {\bf maximun mixing} and a mass-squared difference in the milli-e$V^2$ range:
\begin{equation}
 \sin^2 2\theta \simeq 1~~;~~~~~\Delta m^2 \simeq 3\times 10^{-3} {\rm eV^2}.
\end{equation}

The SuperKamiokande evidence for neutrino masses
already has important implications since it
gives a {\bf lower bound} on 
some neutrino mass:  $m_3\geq \sqrt{\Delta m^2 }\geq 5\times 10^{-2}~{\rm eV}$.  This mass value, in turn, gives a lower bound for the cosmological
contribution of neutrinos to the Universe's energy density:
\begin{equation}
\Omega_\nu \geq \frac{m_3}{92~{\rm eV}~h^2} \sim 1.5\times 10^{-3}~.
\end{equation}
Although this number is far from that needed for closure of the Universe, the  contribution of neutrinos to $\Omega$ 
is comparable to that of luminous matter, $\Omega_{\rm luminous}\sim (3-7)\times 10^{-3}~$.
~\cite{luminous}

For particle physics, a value of $m_3\sim 5\times 10^{-2}~{\rm eV}$ is also
quite interesting, since it provides the best evidence for "new physics" to date. Because neutrinos are neutral, one can write different type of mass terms for them:~\cite{RDP} 

\begin{eqnarray}
{\cal{L}}_{\rm mass}^\nu = &-&[\overline{\nu_{\rm R}}m_D\nu_{\rm L} +
\overline{\nu_{\rm L}}m_D^\dagger\nu_{\rm R}] - \frac{1}{2}
[\overline{\nu_{\rm R}}\tilde C m_S\overline{\nu_{\rm R}}^T + \nu^T_{\rm R}
\tilde Cm^\dagger_S\nu_{\rm R}] \nonumber \\
&-& \frac{1}{2}[\nu_{\rm L}^T\tilde Cm_T\nu_{\rm L} +
\overline{\nu_{\rm L}}\tilde C m_T^\dagger\overline{\nu_{\rm L}}^T]~.
\end{eqnarray}
Here $\tilde{C}$ is a charge conjugation matrix and the mass matrices $m_D,m_S,m_T$ are Lorentz scalars.
However, their presence is only possible as a result of different symmetry
breakdowns.  Specifically, $m_D$, often called a Dirac mass, conserves fermion number, but violates
$SU(2)\times U(1)$ since it does not transform as an $SU(2)$ doublet.  Clearly $m_D$ is proportional to the Fermi scale $v_F$ and is similar to the mass terms that gets induced for quarks and charged leptons, after the electroweak breakdown. Thus the eigenvalues of this matrix should be of the same order as those of the quarks and charged leptons.  Both $m_S$ and $m_T$ violate fermion number by two units and are known
as Majorana masses.  Because $m_S$ couples $\nu_{\rm R}$ with itself,
clearly it is an $SU(2)\times U(1)$ invariant. Thus the eigenstates of $m_S$ are totally unconstrained and are new parameters in the theory. This is not the case for
$m_T$, which violates $SU(2)\times U(1)$ because it does not transform as
an $SU(2)$ triplet. Naively, because of its transformation law under $SU(2)\times U(1)$, one would expect the eigenvalues of this matrix to scale as
$v_F^2/\Lambda$, with $\Lambda$ again being a new parameter in the theory.

Because the  neutrino masses inferred from the SuperKamiokande experiment are in the sub-eV range, and hence much less than $m_\ell$ and $m_q$, it is clear that the  neutrino Majorana mass terms must play a role. Hence one learns that not only individual lepton number, but also total lepton number must be violated. If one assumes that there are no right handed neutrinos, then the neutrino mass matrix is only $m_T$ and one can write for $m_3$ the formula $m_3\sim v_F^2/\Lambda.$
Using the SuperKamiokande result, $\Lambda$ is clearly a very high scale: $\Lambda\sim 10^{15} $ GeV--- a scale of the order of the GUT scale! 

Alternatively, if $\nu_R$ exists (and one neglects $m_T$), then the neutrino mass matrix reads simply~\cite{YGRS}
\begin{equation}
M = \left(
\begin{array}{cc}
0 & m_D^T \\ m_D & m_S 
\end{array} \right)~.
\end{equation}
If the eigenvalues of $m_S$ are large, then M has
two eigenmatrices, given approximately by $m_S$ and $-m_D^Tm_D/m_S$. 
In this case, the spectrum splits into a very heavy neutrino sector and a very light neutrino sector.  This, so called, {\bf see-saw mechanism} is very
suggestive.  For any neutrino, it is natural to expect that the eigenvalues of $m_D$ should be of the order of
the corresponding charged lepton mass. Hence one expects $m_3\sim m_{\tau}^2/m_S$. Again, to fit the SuperKamiokande result requires that there be a large mass scale, now associated with the right-handed neutrinos: $m_S \sim 10^{11}$ GeV.
 
These considerations clearly point to new physics at very large scales, of order of  $10^{11}-10^{15}$ GeV, associated with broken lepton number. This has suggested alternative scenarios for establishing the matter-antimatter asymmetry in the Universe.~\cite{FY} This asymmetry  could originate from a primordial Lepton asymmetry established at temperatures of the order of the scale of Lepton number violation. Since (B+L) processes eventually come into equilibrium in the early Universe, this primordial Lepton asymmetry can get transmuted into a Baryon asymmetry. Remarkably, in this scenario, the CP-violating phases in the neutrino sector are the root cause of our existence! 

\section{Concluding Remarks}
 
Eventhough the SM continues to give an excellent description of the precision electroweak data, the question of the origin of the Fermi scale argues for new physics at, or below, the TeV scale. The most likely form of this new physics, in the view of many, is the presence of an approximate supersymmetry, with spartners which could be as light as 100 GeV. ~\footnote{This is certainly the expected mass range for the lightest Higgs state, for which there are rather reliable bounds.} Perhaps most interestingly, such a low energy supersymmetry has important implications for cosmology. It provides both an interesting candidate for dark matter in the LSP and it may actually make possible baryogenesis at the electroweak scale.

That there is physics beyond the SM has been made clear by the observation of neutrino oscillations. However, this new physics appears to be associated with scales much above the Fermi scale, involving the breakdown of Lepton number. Nevertheless, these phenomena could also have important implications for cosmology. At any rate they indicate already that neutrinos, although they probably do not dominate the Universe's energy density, give a non negligible contribution to this density.

\section*{Acknowledgements}

This work was supported in part by the Department of Energy under contract No. DE-FG03-91ER40662, Task C.

\end{document}